\documentclass[twocolumn,preprintnumbers,amsmath,amssymb]{revtex4}
\usepackage{graphicx}
\usepackage{dcolumn}
\usepackage{bm}
\usepackage{dsfont}
\usepackage{color}
\begin{document}

\title{Effect of an electromagnetic field on the spectra and elliptic flow of particles}
\author{Bohao Feng, Zeyan Wang}
\affiliation{Physics Department, Tsinghua University, Beijing
100084, China}

\begin{abstract}
In 2+1 dimensions, the evolution of flow under the influence of an external electromagnetic field is simulated. The external electromagnetic field is exponentially decaying with time. Under the same initial conditions, flow evolution with and without the external electromagnetic field is compared. It was found that the production of particles was enhanced when the external electromagnetic field was present. As the strength of the electromagnetic field increased, more particles  were produced. Enhancement is greater at the high transverse momentum  than at the low transverse momentum. In addition, the electromagnetic field also modifies the elliptic flow and the modification is related to the mass and charge of the particles.
\end{abstract}
\maketitle

\section{Introduction}
Heavy ion collision is a key mechanism which is utilised when investigating the properties of hot and dense strongly interacting matter. Data obtained from the Relativistic Heavy Ion Collider (RHIC) and the Large Hadron Collider(LHC) indicate that a deconfined quark-gluon plasma (QGP) is created in the collisions \cite{Xu:2014jsa,Adams:2005dq,Adcox:2004mh}. Characterizing and understanding the properties of the QGP and its hadronization is the ultimate goal of heavy-ion collision research. Time-evolution of the QGP can be modeled with relativistic fluid dynamics. Hydrodynamic models have been very successful in providing a quantitative explanation for the behavior observed in experiments \cite{Heinz:2005bw,Shen:2014vra}. However, most of the previous models did not take into account the effects of an electromagnetic field. In reality, the electromagnetic effect is of paramount importance in the heavy ion collision as non-central collisions will produce strong transient electromagnetic fields. Some studies have shown that $e|\vec{B}|/m^2_\pi\approx1-3$  after collisions in the RHIC and $e|\vec{B}|/m^2_\pi\approx10-15$  after LHC collisions \cite{Kharzeev:2007jp,Tuchin:2010vs,Deng:2012pc}  . If electric conductivity is taken into account, the magnetic field will survive even longer.

In recent years, there have been many studies on the consequences of the presence of an electromagnetic field during collisions. In particular, the presence of a magnetic field has led to a lot of interesting phenomena, such as the chiral magnetic effect \cite{Kharzeev:2007jp,Fukushima:2008xe,Manuel:2015zpa} and the Chiral Separation Effect \cite{Kharzeev:2010gd,Burnier:2011bf}.

This paper is organized as follows: In Sec.II the hydrodynamics equations in the presence of an external electromagnetic field in (2+1) dimensions are presented. With an external electromagnetic field, the distribution function of particles is non-equilibrium. The deviations from the local equilibrium, induced by the electromagnetic field, are also introduced in this section. Transverse momentum spectra and the elliptic flow of the particles are the important observables in heavy ion collisions. Thus, the effects of the electromagnetic field on the spectra and elliptic flow path are investigated and presented in Sec.III. Calculations are made in a (2+1)-dimensional hydrodynamic model and the numerical results are presented in this section also. Finally, in Sec.IV, summary and conclusions are given.
\section{(2+1)-dimension hydrodynamics in an external electromagnetic field }
\subsection{Basics of (2+1)-dimension hydrodynamics in an external electromagnetic field }
According to Ref.\cite{Huang:2009ue}, the electromagnetic tensor $F^{\mu\nu}$ can be written as,
\begin{eqnarray}
F^{\mu\nu}
= F^{\mu\lambda}u_\lambda u^\nu- F^{\nu\lambda}u_\lambda u^\mu+\Delta^\mu_\alpha F^{\alpha\beta}\Delta^\nu_\beta \nonumber\\
= E^\mu u^\nu- E ^\nu u^\mu+\frac{1}{2}\epsilon^{\mu\nu\alpha\beta}(u_\alpha B_\beta -u_\beta B_\alpha)
\end{eqnarray}
where $E^\mu\equiv F^{\mu\nu}u_\nu$ and $B^{\mu}=\epsilon^{\mu\nu\alpha\beta}F_{\nu\alpha}u_\beta/2$ are the electric and magnetic fields respectively. Here $\epsilon^{\mu\nu\alpha\beta}$ denotes the anti-symmetric Levi-Civita tensor.

The electromagnetic field obeys Maxwell's equations\cite{DelZanna:2007pk}
\begin{eqnarray}
d_\mu F^{\mu\nu}=J^{\nu},~d_\mu J^{\mu}=0
\end{eqnarray}
where $J_{\nu}$ is the current density and $d_\mu$ is the covariant derivative. The relationship between the current and the fields is assumed to take the linear form\cite{ Inghirami:2016iru}
\begin{eqnarray}
J^{\mu}=\rho_e u^\mu+j^\mu , ~j^\mu=\sigma E^\mu
\end{eqnarray}
where $\rho_e$ is the electric charge density,$j^\mu$ is the conduction current and $\sigma$ is the plasma conductivity tensor (which is assumed to be constant in the calculations presented in this paper).The energy-momentum tensor $T^{\mu\nu}$ satisfies\cite{ Inghirami:2016iru,DelZanna:2007pk},
\begin{eqnarray}
d_\mu T^{\mu\nu}=J_{\nu}F^{\mu\nu}
\end{eqnarray}

In order to solve Eq.(4), the initial values and time evolutions of the external electromagnetic field must be specified in advance. According to Ref.\cite{Deng:2012pc}, it is known that the longitudinal fields $B_z, E_z$ are much smaller than the transverse fields $B_{x,y}$ and $E_{x,y}$. Therefore, $B_z, E_z$ are neglected in the calculations presented here. In addition, the possible polarization and magnetization effects of the plasma are also neglected.

 Details of time and space dependence of the magnetic fields can be found in Refs.\cite{Deng:2012pc,Tuchin:2013apa,Hirono:2016lgg}.In the present demonstrative calculations, the spatial distributions of the magnetic fields is parametrized as follows\cite{Hirono:2016lgg}:
\begin{eqnarray}
eB_y(\vec{r},t)=eB_x(\vec{r},t)=eB_0\exp(-\frac{r^2}{2\sigma^2_{r}}-\frac{t}{\tau_B})
\end{eqnarray}

The electric field can be calculated (Ref.\cite{Deng:2012pc}) by $e\textbf{\textit{E}}=-\textbf{\textit{v}}\times e\textbf{\textit{B}}$. Here $r =\sqrt{ x^2+ y^2 }$ and $\sigma_{r} = 2.0 fm$. The initial value of the field is taken to be $ eB_0 = 0.08 GeV^2 \simeq(2m_\pi)^2 $, which is estimated in Ref.\cite{Bzdak:2011yy}.The duration time of  magnetic field is controlled by the parameter $\tau_B$ (see Ref.\cite{Tuchin:2013apa,Hirono:2016lgg}).

In (2+1)-dimension hydrodynamics, heavy ion collisions are best described in the coordinates $(\tau,x,y,\eta)$, which are related with the flat coordinates $(t,x,y,z)$. The longitudinal proper time and the rapidity are:
\begin{eqnarray}
\tau=\sqrt{t^2-z^2},~~\eta=\frac{1}{2}\ln\frac{t+z}{t-z}
\end{eqnarray}

The metric tensor associated with the $(\tau,x,y,\eta)$ coordinates is
\begin{eqnarray}
g^{\mu\nu}=diag(1,-1,-1,-1/{\tau}^2)
\end{eqnarray}

This leads to the following non-vanishing Christoffel symbols
\begin{eqnarray}
\Gamma^\eta_{\eta\tau}=\Gamma^\eta_{\tau\eta}=\frac{1}{\tau},\Gamma^\tau_{\eta\eta}=\tau
\end{eqnarray}

In this coordinate system, conservation laws for the energy momentum tensor $T^{\mu\nu}$ can be written as (see Ref.\cite{Kolb:2000sd}):
\begin{eqnarray}
\partial_\tau\widetilde{T}^{\tau\tau}+\partial_x(v_x\widetilde{T}^{\tau\tau})+\partial_y(v_y\widetilde{T}^{\tau\tau})\nonumber\\
=-p-\tau \partial_x(p v_x)-\tau \partial_y(p v_y)+J_{x}F^{x\tau}+J_{y}F^{y\tau}\nonumber\\
\partial_\tau\widetilde{T}^{\tau x}+\partial_x(v_x\widetilde{T}^{\tau x})+\partial_y(v_y\widetilde{T}^{\tau x})\nonumber\\=
-\tau \partial_x(p )+J_{y}F^{y x}+J_{\tau}F^{\tau x}\nonumber\\
\partial_\tau\widetilde{T}^{\tau y}+\partial_x(v_x\widetilde{T}^{\tau y})+\partial_y(v_y\widetilde{T}^{\tau y})\nonumber\\=
-\tau \partial_y(p )+J_{x}F^{x y}+J_{\tau}F^{\tau y }
\end{eqnarray}
Here $\widetilde{T}^{\mu\nu}=\tau T^{\mu\nu}$and $v_i$ are the velocity components of the fluid.

 In contrast to the ideal case, conservation equations now contain the electromagnetic tensor $F^{\mu\nu}$ and current density $J^\mu$ in the external magnetic field. Therefore, the components of the energy-momentum tensor will evolve under the influence of the electromagnetic field. To solve the differential equations (9), the ``Sharp and Smooth Transport Algorithm" (SHASTA\cite{Boris,Rischke:1995ir})method  was employed. Given an equation of state, if the energy density ($\varepsilon$) and fluid velocity ($v_x$,$v_y$ ) distributions at time $\tau_i$ are known, $\varepsilon$,$v_x$ and $v_y$ at the next time step $\tau_{i+1}$ can be obtained from differential equations(9).

\subsection{The correction of the distribution function }
With an electromagnetic field present, the system is not in equilibrium.In order to correct the distribution function, the same method that was employed in Ref. \cite{Greif:2014oia} is used.
The time evolution equation satisfied by $f(x,p)$ is the Boltzmann equation
\begin{eqnarray}
p^\mu\frac{\partial}{\partial x^\mu}f+q_e p_\nu F^{\mu\nu}\frac{\partial}{\partial p^\mu}f=\sum C_{ij}(x^\mu,p^\mu)
\end{eqnarray}

Here $C_{ij}$ is the collision term and $q_e$ is the electric charge of the particle.$F^{\mu\nu}$ is the electromagnetic tensor,as defined by Eq.(1). It is assumed that the distribution function which satisfies Boltzmann equations (10) is $f=f_0+\delta f$, where $f_0$ is the equilibrium distribution function (Bose-Einstein or Fermi-Dirac)
\begin{eqnarray}
f_0=1/ (\exp(u_\mu p^\mu/T)\pm 1)
\end{eqnarray}

In the case of the relaxation time limit, the Boltzmann equation can be rewritten as (see Ref. \cite{Greif:2014oia} for details):
\begin{eqnarray}
p^\mu\frac{\partial}{\partial x^\mu}f+q_e p_\nu F^{\mu\nu}\frac{\partial}{\partial p^\mu}f=-\frac{p^\mu u_\mu}{\tau_{rel}}(f-f_0)
\end{eqnarray}
and in local rest frame,we have
\begin{eqnarray}
\delta f=\frac{1}{T}\frac{q_e \tau_{rel}}{p^\mu u_\mu}f_0(1\pm f_0) p_{\nu} F^\nu
\end{eqnarray}
where $F^{v}=(E^v-\frac{1}{2}u_\mu\epsilon^{\mu\nu\alpha\beta}( u_\alpha B_\beta-u_\beta B_\alpha))$. In the calculations presented here $ \tau_{rel}=1.0 fm/c$.

It should be mentioned that the correction to the distribution function can also be derived by using the method outlined in Ref.\cite{Arnold:2000dr}. The form of the correction can be written as
\begin{eqnarray}
\delta f=f_0(1\pm f_0)\frac{q_e \hat{p}_v F^{v}}{T^{2}}\chi(p/T)
\end{eqnarray}
where $\chi(p/T)=(|p|/T)$,$\hat{p}_v=p_v/|p^\mu u_\mu|$ is the momentum unit vector.

\subsection{Initial conditions}
In the present calculations, similar initial conditions as Ref. \cite{Shen:2014vra} have been used. For simplicity and comparison with the ideal case a simple Glauber model is used. Following Ref. \cite{Shen:2014vra}, initial transverse energy density  distribution is given by the Glauber model.The collision at impact parameter $\vec{b}$ involves 80\% soft scattering and 20\% hard scattering.

In order to fit the charge multiplicity for most central collisions at RHIC  and  LHC ,we set $\varepsilon_0=55GeV/fm^3$ at RHIC and $\varepsilon_0=166GeV/fm^3$at LHC at the initial time $\tau_0$=0.4 fm. In addition, it is assumed that at the initial time $\tau_0$=0.4 fm, the initial transverse flow velocities are zero, $v_x(x,y) = v_y (x,y)=0$,and the shear viscosity $\eta/s=0$ . The parameter $\varepsilon_0$ and the initial equilibration time $\tau_0$ are fixed to reproduce the experimental results of the heavy ion collisions.

\section{Results and Discussions}
\subsection{Electromagnetic effect on the spectra and elliptic flow of hadrons}

It is assumed that when the temperature drops below the critical value, non-interacting particles instantaneously freeze-out from a thermalize fluid to free-streaming. The hadron spectra can be obtained by using the Cooper-Frye freezeout prescription (see Ref.\cite{Cooper:1974mv}) over the freeze-out surface $\Sigma(x)$,
\begin{eqnarray}
E\frac{d^3 N}{d^3p}&=&\frac{g}{(2\pi)^3}\int p\cdot d^3\sigma(x)f(x,p)\nonumber\\
&=& \frac{g}{(2\pi)^3}\int p\cdot d^3\sigma(x)[f_0(x,p)+\delta f(x,p)]
\end{eqnarray}
 where $g$ is the degeneracy factor for particles, $p^\mu$ is the particle four momentum.The surface $\Sigma(x)$ is subdivided into infinitesimal elements $d^3\sigma(x)$ and $d^3\sigma_\mu(x)$is the outward-pointing vector with the magnitude $d^3\sigma(x)$ on the surface $\Sigma(x)$ at point $x$ \cite{Heinz:2004qz},
\begin{eqnarray}
d^3\sigma_\mu=(\cosh \eta,-\frac{\partial \tau_f }{\partial x_f},-\frac{\partial \tau_f}{\partial y_f},-\sinh \eta )\tau_f r dr d\phi d\eta
\end{eqnarray}

The integration measure in the Cooper-Frye formula (15) is
\begin{eqnarray}
p\cdot d^3\sigma(x)=[m_T\cosh(y-\eta)-\textbf{p}_\perp \nabla _\perp \tau_f]\tau_f r dr d\phi d\eta
\end{eqnarray}

Here $\tau_f$ is a function of the transverse coordinate $\textbf{r}$,and $m_T=\sqrt{p^2_T+m^2}$ is the particle's transverse mass.

In Eq.(15), the particle spectrum is a sum of a local equilibrium and a non-equilibrium contribution. The equilibrium distribution function $f_0(x,p)$ and the deviation from equilibrium distribution function $\delta f(x,p)$ are given in the previous section. The flow velocity in the distribution $u^\mu =\gamma_\perp (\cosh\eta,v_\perp \cos\varphi_v ,v_\perp \sin\varphi_v ,\sinh\eta)$ with $\gamma_\perp=\frac{1}{\sqrt{1-v^2_x-v^2_y}}$ is parameterized along the surface $\Sigma(x)$ and the scalar product $p^\mu u_\mu$ in the distribution then becomes
\begin{eqnarray}
p^\mu u_\mu=\gamma_\perp[m_T\cosh(y-\eta)-p_T v_\perp\cos(\phi_p-\phi_v)]
\end{eqnarray}

The azimuthal anisotropy of the particle momentum distribution can be characterized by $v_n$, the coefficients of Fourier expansion decomposition with respect to the azimuthal angle $\phi_p$ \cite{Voloshin:1994mz,Luzum:2013yya}:
\begin{eqnarray}
E\frac{dN^3}{d^3p}&=&\frac{dN}{dy p_T d p_T d\phi_p}\nonumber\\
&=&\frac{1}{2\pi}\frac{d N}{p_T d p_T dy}[1+2\sum v_n\cos n(\phi_p)]
\end{eqnarray}

Once the spectrum has been determined, the anisotropic flow coefficients can be calculated and elliptic flow $v_2$ defined as,
\begin{eqnarray}
v_2(p_T)=\frac{\int^{2\pi}_0\frac{d N}{dy p_T d p_T d\phi_p}\cos(2\phi_p)d\phi_p}{\int^{2\pi}_0\frac{d N}{dy p_T d p_T d\phi_p}d\phi_p}
\end{eqnarray}

In Fig.1, the spectra of $\pi^+$ and $K^+$ obtained in the Cooper-Frye formalism are shown.As in the ideal fluid case, freeze-out temperature $T_{F} =130MeV$.By comparing the numerical results from the hydrodynamic simulations with experimental data measured,we found that over the full $p_T$ range, the deviations between the numerical results and data measured  are all within 15\% ,no matter whether the external electromagnetic field is present or not.For spectra of $\pi^+$,the particle production from ideal hydrodynamic simulation without electromagnetic field are about 10\% lower than with electromagnetic field when transverse momentum $p_T>2.5GeV$.It indicates that particle production will increase due to the electromagnetic corrections and the effect of the electromagnetic field is more prominent at large $p_T$ than at low $p_T$.The electromagnetic field influences particle production in two ways. On one hand is the impact to the evolution of the flow, on the other hand is the introduction of a correction to the equilibrium distribution function. Non-equilibrium contribution depends on the strength of the electromagnetic field. Therefore, the spectra for hadrons obtained with or without an electromagnetic field do not differ very much in the present calculations. This is a consequence of the rapid weakening of the electromagnetic field over time , as described by Eq.(5).
\begin{figure}
\includegraphics[width=8.6cm]{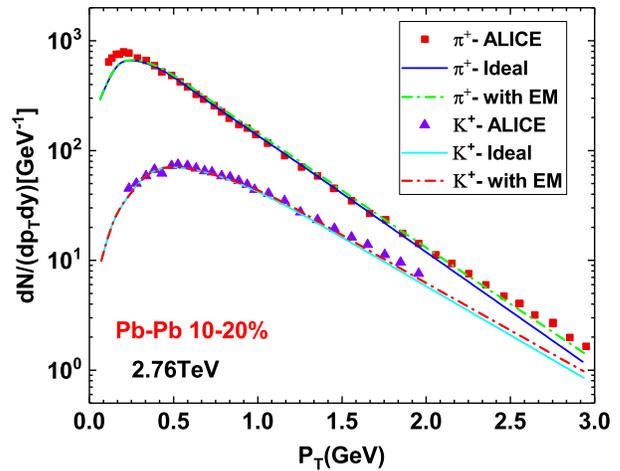}
\caption{\label{fig1} Transverse momentum spectra of $\pi^+$and $K^+$ in Pb-Pb collisions in centralities 10-20\%, obtained in the ideal hydrodynamic
model with (dash-dotted lines)and without an external electromagnetic field (solid lines). Symbols represent data of the ALICE Collaboration \cite{Floris:2011ru}}
\end{figure}

In Fig.2 the momentum-dependent elliptic flow coefficients for $\pi^+$ and $K^+$ are shown. The elliptic flow for the ideal case is compared to the case with the electromagnetic field. The electromagnetic field acts on the evolution of flow, leading to the reduction of elliptic flow of positively charged particles. Whether or not there is an electromagnetic field, the change of elliptic flow of hadrons is not great because the electromagnetic field is weak in the final stage of the expansion. It was also found that the non-equilibrium correction stemming from the electromagnetic field could in principle be very different for different species of particles, as it depends on the mass and charge of the particles.

It is worth mentioning that all of the hydrodynamic results in Fig. 2 do not agree well with the experimental data at higher $p_T$ . The reason is that the shear viscosity and bulk viscosity were not considered in the calculations presented in this paper. The elliptic flow $v_2$ of hadrons will be suppressed at higher $p_T$ by both shear and bulk viscosity corrections.
\begin{figure}
\includegraphics[width=8.6cm]{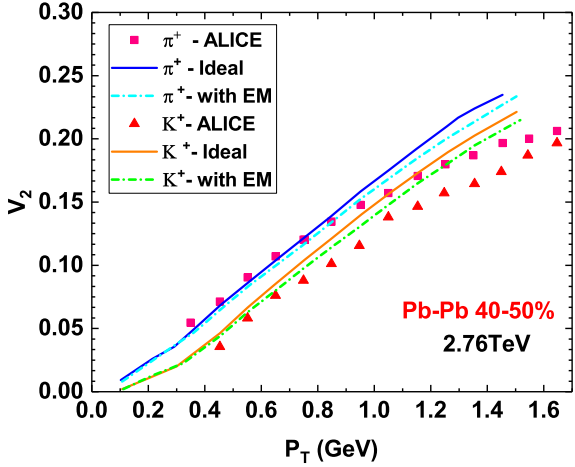}
\caption{\label{fig2} Elliptic flow coefficients of the identified particles as a function of transverse momentum in the ideal case (solid lines)or with an electromagnetic field (dash-dotted lines); the symbols represent the results of the ALICE Collaboration \cite{Krzewicki:2011ee}, centrality 40-50\%.}
\end{figure}

\subsection{Electromagnetic effect on the spectrum and $v_2$ of photons}
In relativistic heavy ion collisions, photons are powerful probes for QGP and magnetic field has impact on phenomenology of QGP \cite{Tuchin:2011jw,Mohapatra:2011ph}.

When photon comes out of the dense and strongly-interacting medium and is detected, most of the information it carries is not lost.The transverse-momentum spectra and azimuthal asymmetry of the photons can reflect the properties of the substance under extreme conditions.It is essential to investigate the effect the electromagnetic field has on the spectrum and elliptic flow of the photons\cite{Tuchin:2012mf,Bzdak:2012fr,Yee:2013qma,Muller:2013ila}.

In order to obtain the photon spectrum,we have to calculate the photon production rate first. The photon production rate can be computed within a kinetic description.For $2\rightarrow 2$ photon production process($1+2\rightarrow3+\gamma$),it is\cite{Dion:2011pp}
 \begin{eqnarray}
k\frac{d^3R^\gamma}{d^3k}&=&\frac{1}{2(2\pi)^3}\int\frac{d^3p_1}{2E_{P_1}(2\pi)^3}\frac{d^3p_2}{2E_{P_2}(2\pi)^3}\frac{d^3p_3}{2E_{P_3}(2\pi)^3}\nonumber\\
&\times&(2\pi)^4\delta^{(4)}(P_1+P_2-P_3-K)|M|^2f_{p_1}f_{p_2}\nonumber\\
&\times&(1\pm f_{p_3})
 \end{eqnarray}
where$|M|^2$ is the squared matrix element corresponding to the photon emission process.$f_p$ is the distribution function for bosons and fermions.It is modified in the presence of magnetic field.According to the previous section, the modification can be written as $f_p=f^{(0)}_p+\delta f$.Here$f^{(0)}_p$ is correspondingly either the Fermi-Dirac or Bose-Einstein distribution.

Inserting the modified distribution function into Eq.(21),one may have
 \begin{eqnarray}
 k\frac{d^3R^\gamma}{d^3k}=k\frac{d^3R^{\gamma(0)}}{d^3k}+k\frac{d^3\delta R^{\gamma(EM)}}{d^3k}
 \end{eqnarray}
 where
  \begin{eqnarray}
  k\frac{d^3\delta R^{\gamma(EM)}}{d^3k}&=&\frac{1}{2(2\pi)^3}\int\frac{d^3p_1}{2E_{P_1}(2\pi)^3}\frac{d^3p_2}{2E_{P_2}(2\pi)^3}\frac{d^3p_3}{2E_{P_3}(2\pi)^3}\nonumber\\
&\times&(2\pi)^4\delta^{(4)}(P_1+P_2-P_3-K)|M|^2\nonumber\\
&\times& [\delta f_{p_1}f^{(0)}_{p_2}(1\pm f^{(0)}_{p_3})+f^{(0)}_{p_1}\delta f_{p_2}(1\pm f^{(0)}_{p_3})\nonumber\\
&\pm & f^{(0)}_{p_1}f^{(0)}_{p_2}\delta f_{p_3}]
 \end{eqnarray}

Photon  spectra can be obtained by integrating the photon emission rate over space-time\cite{Bhattacharya:2015ada}
 \begin{eqnarray}
E\frac{d^{3}N^\gamma}{d^3p}=\frac{dN^\gamma}{d\phi_p p_Tdp_Tdy}=\int dx^4 \frac{dR^\gamma}{d\phi_p p_Tdp_Tdy}
 \end{eqnarray}

Now we turn to the elliptic flow of direct photons.Fourier decomposition of the differential spectrum was made  to find out the dependence of photon production on the azimuthal angle.In a particular event,Fourier coefficient $v^s_n$and event-plane angle $\Psi^s_n$ are given by\cite{Vujanovic:2016anq,Shen:2015qba}
\begin{eqnarray}
v^s_ne^{in\Psi^s_n}=\frac{\int dp_Tdyd\phi p_T[p^0\frac{d^3N^s}{d^3p}]e^{in\phi}}{\int dp_Tdyd\phi p_T[p^0\frac{d^3N^s}{d^3p}]}
\end{eqnarray}
where superscript ``s" represents the particle species.

Generally,the anisotropic flow coefficients of direct photons are calculated by correlating them with charged hadrons\cite{Shen:2015qba}.Using the scalar-product method,one may obtain the anisotropic flow of direct photons as \cite{Vujanovic:2016anq,Shen:2015qba,Shen:2014lpa}:
 \begin{eqnarray}
 v^\gamma_n\{SP\}=\frac{<v^h_nv^\gamma_n \cos[n(\Psi^\gamma_n-\Psi^h_n)]>_{ev}}{\sqrt{<(v^h_n)^2>_{ev}}}
 \end{eqnarray}
where $<...>_{ev}$ is an average over events.In present calculation,we set $n=2$ to obtain the elliptic flow of photons.
\begin{figure}
\includegraphics[width=8.6cm]{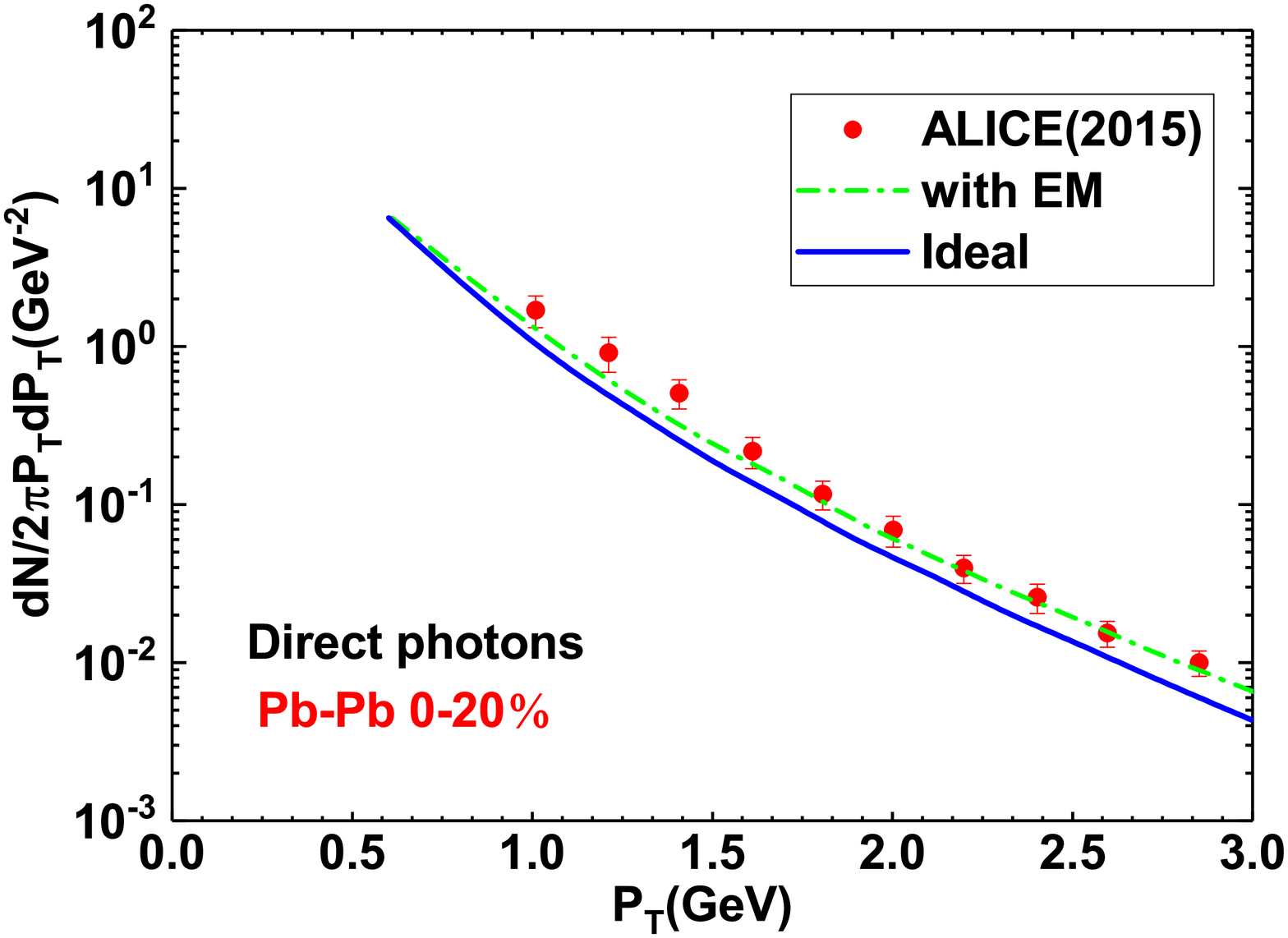}
\caption{\label{fig3}  Transverse momentum spectra of photons in Pb - Pb collisions at the LHC in the 0 - 20\% centrality range. The solid curve represents the ideal case, while the dash-dotted line represents the case in the electromagnetic field. The data is from Ref.\cite{Adam:2015lda}}
\end{figure}

The photon spectra with and without the electromagnetic field at $\sqrt{s_{NN}}$ = 2760 GeV in Pb-Pb collisions are presented in Fig.3. By comparison, it is easy to see that the effect the electromagnetic field has on the spectrum of direct photons is similar to that on the spectra of hadrons, except that the influences are more pronounced. The reason for this is that the photon yield not only originates from the QGP phase but also from the hadronic phase. Compared to the hadronic phase, in the QGP phase the electromagnetic field is stronger. In addition, the effect time of the electromagnetic field on the photon spectrum is longer. The influence of the electromagnetic field on this spectrum is mainly due to its effect on the photon emission rates and space-time evolution of the medium.
\begin{figure}
\includegraphics[width=8.6cm]{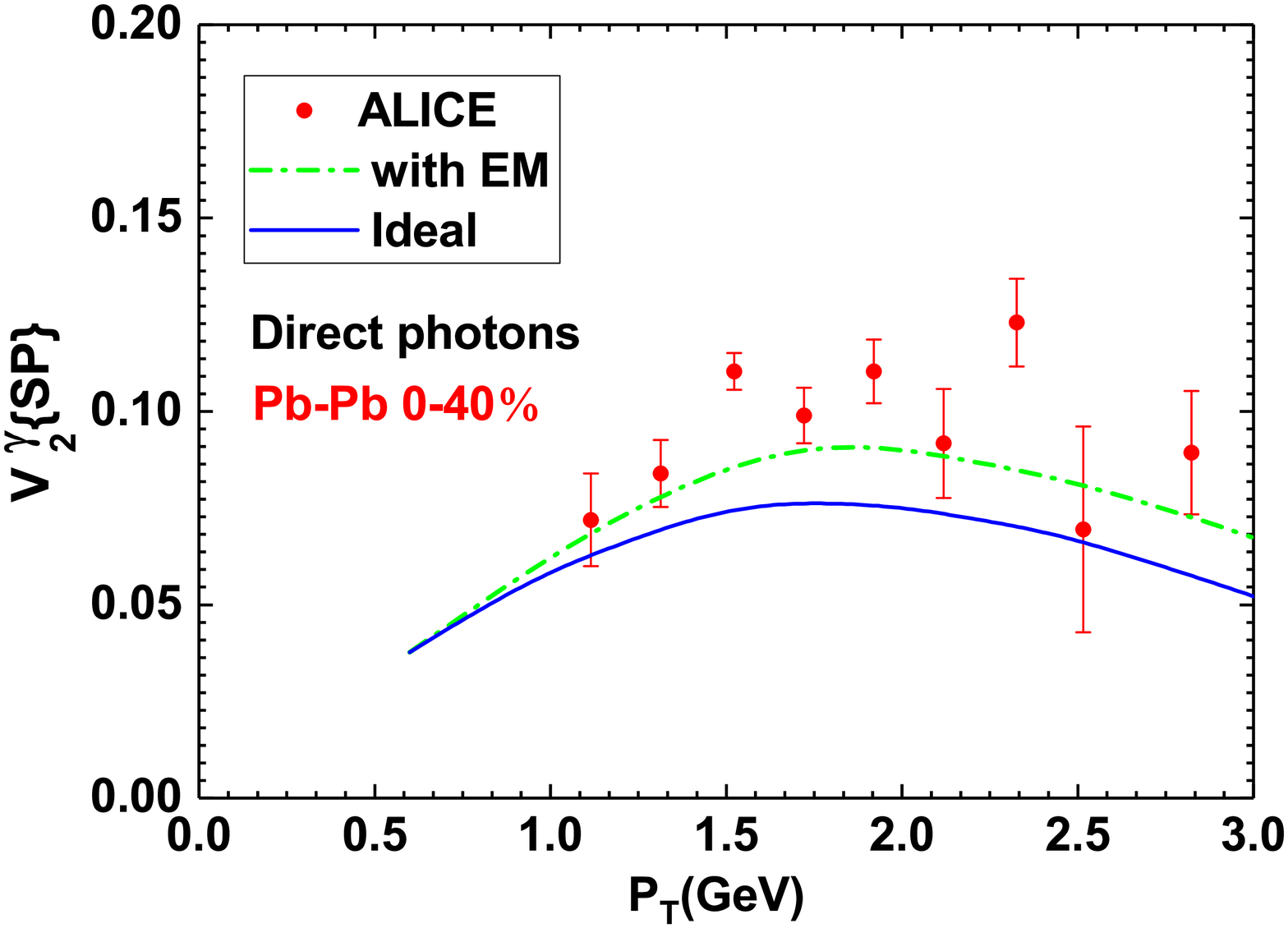}
\caption{\label{fig4} Elliptic flow $v_2$ of the photons as a function of the transverse momentum $p_T$. The solid blue line is for an ideal fluid. The dash-dotted green line is for the fluid with an electromagnetic field. The symbols represent the results of the ALICE Collaboration \cite{Lohner:2012ct,Wilde}, centrality 0-40\%.}
\end{figure}

The elliptic flow $v_2(p_T )$ of photons produced in 0-40\% central Pb+Pb collisions at $\sqrt{s_{NN}}$ = 2760 GeV from the hydrodynamic simulations with and without electromagnetic field (solid color line) are shown in Fig.4. Results are compared to the preliminary data from the ALICE collaboration \cite{Lohner:2012ct,Wilde}. The $v_2$ of the photons increases at low $p_T$ and decreases at high $p_T$. Comparing the results with and without the electromagnetic effect, it was found that adding the electromagnetic field increase the elliptic flow and the differences between the cases with or without an electromagnetic field are larger for photons than for hadrons. One may conclude that the elliptic flow of photons is sensitive to the electromagnetic effect in the early stage.

\section{Summary and Conclusions}
A study of the effect an external electromagnetic field has in relativistic heavy-ion collisions is presented. Its impacts on both hadrons and photons are investigated assuming a small electromagnetic field strength. Both the ideal case and the case with an electromagnetic field are initialized equally and the transverse profile is taken from a Glauber model calculation. The effects of the electromagnetic field on the spectra and elliptic flow of particles are examined. With the electromagnetic field, better agreement was found with preliminary experimental data for the transverse momentum spectra of pions, kaons and photons. The electromagnetic field leads to enhance of the particle production. Particle production increase due to non-equilibrium correction to the equilibrium distribution function and the effect of the electromagnetic field on the evolution of medium. Non-equilibrium correction to the equilibrium distribution function is a dominating factor which influences the particle production. It is also noted that the effect of the electromagnetic field is more prominent at large $p_T$ than at low $p_T$. The electromagnetic field not only enhances particle production, but it also affects the elliptic flow. The elliptic flow coefficient of hadrons is only slightly affected by the electromagnetic field, however, the effect of the electromagnetic field on the elliptic flow coefficient of the photons is significant. In other words, the elliptic flow of photons is more sensitive to electromagnetic effects. The main reason for this is that the electromagnetic field rapidly weakens over time,and thus its effect is stronger in the QGP phase than the hadronic phase.

To conclude, the present study shows that the inclusion of the electromagnetic field in the medium is very important in the analysis of results obtained from RHIC and LHC collisions. A more accurate description of transverse momentum spectra and elliptic flow of particles is achieved when it is included.

\vspace{3mm}
\begin{acknowledgments}
We would like to thank K.Zhou,M.Greif and C.Greiner at Goethe University Frankfurt for helpful discussions.This work was financially supported by the NSFC and the MOST under Grants No.11275103 and No.2015CB856903.
\end{acknowledgments}

\end{document}